\documentclass[10pt, oneside]{article}   	% use "amsart" instead of "article" for AMSLaTeX format

\usepackage{geometry}          
\usepackage{graphicx}
\usepackage{mathtools}		
\usepackage{amssymb}
\usepackage{amsthm}
\usepackage{amsmath}
\usepackage{amsfonts}

\usepackage{subcaption}
	
\usepackage{setspace} 
\usepackage[utf8]{inputenc}
\usepackage[T1]{fontenc}
\usepackage{authblk}

\usepackage[printonlyused]{acronym}

\acrodef{lopa}[LOPA]{Layers of Protection Analysis}
\acrodef{bdb}[BDB]{Beyond Design Basis}
\acrodef{uq}[UQ]{Uncertainty Quantification}

\acrodef{cap}[CAP]{Condition Adverse to Quality Program}
\acrodef{cba}[CBA]{Cost-Benefit Analysis}
\acrodefplural{cba}[CBAs]{Cost-Benefit Analyses}
\acrodef{cdf}[CDF]{Core Damage Frequency}
\acrodef{chrs}[CHRS]{Containment Heat Removal System}
\acrodef{crmp}[CRMP]{Comprehensive Risk Management Process}
\acrodefplural{crmp}[CRMPs]{Comprehensive Risk Management Processes}
\acrodef{db}[DB]{Design-basis}
\acrodef{dba}[DBA]{Design Basis Accident}
\acrodef{did}[DID]{Defense-in-Depth}
\acrodef{eccs}[ECCS]{Emergency Core Cooling System}
\acrodef{esd}[ESD]{Event Sequence Diagram}
\acrodef{flc}[FC]{Fails Closed}
\acrodef{fda}[FDA]{Food and Drug Administration}
\acrodef{fmea}[FMEA]{Failure Mode and Effects Analysis}
\acrodefplural{fmea}[FMEA]{Failure Modes and Effects Analyses}
\acrodef{fsar}[FSAR]{Final Safety Analysis Report}
\acrodef{fs}[FoS]{Factor of Safety}
\acrodefplural{fs}[FsoS]{Factors of Safety}
\acrodef{ft}[FT]{Fault Tree}
\acrodef{fts}[FTS]{Fail To Start}
\acrodef{ftr}[FTR]{Fail To Run}
\acrodef{gra}[GRA]{Generation Risk Assessment}
\acrodef{lerf}[LERF]{Large Early Release Frequency}
\acrodef{mss}[MSS]{Main Steam System}
\acrodef{mslb}[MSLB]{Main Steam Line Break}
\acrodef{nei}[NEI]{Nuclear Energy Institute} 
\acrodef{npv}[NPV]{Net Present Value}
\acrodef{npp}[NPP]{Nuclear Power Plant}
\acrodef{nrc}[NRC]{Nuclear Regulatory Commission}
\acrodef{oqap}[OQAP]{Operations Quality Assurance Program}
\acrodef{lar}[LAR]{License Amendment Requests}
\acrodef{lb}[LB]{Licensing Basis}
\acrodef{lerf}[LERF]{Large Early Release Frequency}
\acrodef{loca}[LOCA]{Loss of Coolant Accident}
\acrodef{lwr}[LWR]{Light Water Reactor}
\acrodef{om}[O\&M]{Operations and Maintenance}
\acrodef{ora}[ORA]{Organizational Risk Assessment}
\acrodef{pga}[PGA]{Peak Ground Acceleration}
\acrodef{pra}[PRA]{Probabilistic Risk Assessment}
\acrodef{pwr}[PWR]{Pressurized Water Reactor}
\acrodef{psa}[PSA]{Probabilistic Safety Analysis}
\acrodef{rcb}[RCB]{Reactor Containment Building}
\acrodef{rcd}[RCD]{Reactor Core Damage}
\acrodef{rcs}[RCS]{Reactor Coolant System}
\acrodef{rr}[RR]{Radiation Release}
\acrodef{ssc}[SSC]{Systems, Structures, and Components}
\acrodef{stm}[STM]{State Transition Matrix}
\acrodefplural{stm}[STMs]{State Transition Matrices}
\acrodef{ufsar}[UFSAR]{Updated Final Safety Analysis Report}
\acrodef{nrc}[NRC]{Nuclear Regulatory Commission}
\acrodef{pasta}[PASTA]{Poisson Arrivals See Time Averages}
\acrodef{loa}[LOA]{Lack of Anticipation}
\acrodef{wp1}[w.p.1]{with probability 1}
\acrodef{ba}[asm]{by assumption}
\acrodef{aa}[a.a.]{almost all}
\acrodef{as}[a.s.]{almost surely}
\acrodef{iid}[i.i.d]{independent and identically distributed}

\usepackage{natbib}

\newtheorem{corollary}{Corollary}
\newtheorem{definition}{Definition}

\newtheorem{proposition}{Proposition}

\newtheorem{remark}{Remark}

\newtheorem*{features}{Predictive Model Features}

\usepackage[colorlinks = true,
            urlcolor  = blue,
            linkcolor = black,
            citecolor = black]{hyperref}
            
\usepackage{subcaption}
\usepackage{natbib}
\usepackage{cleveref}

\makeatletter
\renewcommand\tagform@[1]{\maketag@@@{\ignorespaces#1\unskip\@@italiccorr}}
\makeatother
\creflabelformat{equation}{#2\textup{#1}#3}

\providecommand{\keywords}[1]
{
  \small	
  \textbf{Keywords---} #1
}

\title{Is Core Damage Frequency and Informative Risk Metric?}
\usepackage{authblk}
\author[]{Martin Wortman\thanks{corresponding author, wortman@haztechrisk.org}}
\author[]{Ernie Kee}
\author[]{Pranav Kannan}

\affil[]{HazTechRisk.Org}

\begin{document}

\maketitle

\begin{abstract}
	\ac{cdf} is a risk metric employed by nuclear regulatory bodies worldwide.
	Numerical values for this metric are required by U.S. regulators, prior to reactor licensing, and reported values can trigger regulatory inspections. 
	\ac{cdf} is reported as a constant, sometimes accompanied by a confidence interval.
	It is well understood that \ac{cdf}  characterizes the arrival rate of a stochastic point process modeling core damage events.
	However, consequences of the assumptions imposed on this stochastic process as a computational necessity are often overlooked.
	Herein, we revisit \ac{cdf} in the context of modern counting processes.
	We will argue that the assumptions required to obtain a numerical value for \ac{cdf} (\emph{e.g.,} with \ac{pra}) are typically unrealistic and lead to an underestimate bias.  We will conclude that:
	\begin{enumerate}
	\item The computation of core damage frequency (CDF) cannot account for the stochastic dependence between reactor protections and arriving initiating events.
	\item In practice, there can be no direct observational data reflecting the propensity for core damage.
	\item CDF calculated via PRA or PSA is optimistically biased.
	\item The optimistic bias of \ac{cdf} should disqualify its use in regulatory oversight.
\end{enumerate}
\end{abstract}


\section{Overview}\label{introduction}
	\ac{cdf} has a long history as a risk metric for the U.S. civilian nuclear fleet.
	It is a data--derived constant intended to provide insight as to the likelihood of core damage events.
	\ac{cdf} is used to report both historical fleet performance and to predict performance of individual reactor units.
	As an historical performance metric, \ac{cdf} is simply the quotient of total industry core damage events and total reactor years of operation.
	For example, see \citeauthor{Ha-Duong:2014rw} for such history up to 2014.
	But, risk is concerned with future operational behaviors.  
	Since any practical reactor will experience at most one core damage event,  `frequency' is an 
	unnatural concept that must be understood within the context of modeling used to support the computation of a predictive \ac{cdf}.

	A large body of literature is devoted to computing estimates of \ac{cdf}; most of this work extends from \cite{rasmussen1981}.  
	The central theme of this corpus focuses on how to analytically connect possible core damage with the status of reactor protections 
	and the arrival of initiating events that, if unmitigated by protections, might exceed to catastrophe. 
	Examples, among many reported in the literature, that employ or extend Rasmussen's work include
 	\cite{ZHANG2021111135} who argue that the cost and benefit of safety enhancements can be in part evaluated against predicted \ac{cdf}.
	\cite{CHO201777} describe how, based on \ac{cdf}, some protections are less important than others.
	\cite{WILLIAMS2018157} describe a method that integrates prediction of conditional probability of core damage events with 
	predicted frequency of core damage.
	\cite{HARAGUCHI2020110433} use a combination of \ac{cdf} and core damage probability in a seismic risk study.

	These typical analyses rely on the Rasmussen characterization of \ac{cdf} as the product of a predicted arrival rate of initiating 
	events (denoted by $\lambda$) with a predicted proportion of initiating events that exceed to core damage (denoted by $p$). 
	However, the Rasmussen approach relies on informal mathematics that treats $p$ as the long--run probability reactor protections 
	are unavailable and thereby adopts an implicit assumption that long--run reactor protection unavailability as seen at the epochs of 
	the randomly arriving initiating events is the same as would be observed at an arbitrary (\emph{i.e.,} non--random) time.
	This assumption leads to computationally tractable estimates of \ac{cdf}.  
	Unfortunately, however, it holds only in special circumstances that typically defy practical justification.  
	We will show that assuming the proportion of initiating events that exceed to core damage is equal to the limiting probability that 
	protections are unavailable leads to an optimistic bias when estimating \ac{cdf}.
	\footnote{
	There is a large literature devoted to bias phenomena that arise with stochastic point and counting process models. Famous well--studied examples include the Inspection Paradox, Feller's Paradox, and Palm Probabilities. A common feature of these bias phenomena is the discrepancy between event probabilities as seen across different classes of stopping times. It should be of no surprise that counting processes used to study \ac{cdf} would exhibit a bias phenomenon.
	}
 	We refer to this assumption as the \citeauthor{rasmussen1981} characterization.

	We will explore practical conditions under which the \citeauthor{rasmussen1981} characterization of \ac{cdf} is deficient, and then gauge the 
	usefulness of \ac{cdf} as a risk metric by addressing the following questions:
	\begin{enumerate}
		\item When does \ac{cdf} exist?
		\item Is it feasible to estimate \ac{cdf}, when it exists?
		\item Do estimates of \ac{cdf} closely approximate its true value?
	\end{enumerate}
 	We approach these questions by first introducing three easily proved propositions and their corollaries in \Cref{predictive_model}. 
	We then conclude with \Cref{conclusions} where we directly address these questions by applying the results of \Cref{predictive_model}.
	Our analyses rely on standard results from the theory of martingales and stochastic counting processes.

	%%%%%%%%%%%%%%%%%%%%%%%%%%%%%%%%%%%%%
	%%%%    PREDICTIVE MODEL.   %%%%%%%%%%%%%%%%%%%
	%%%%%%%%%%%%%%%%%%%%%%%%%%%%%%%%%%%%%
\section{Predictive Model}\label{predictive_model}

	Consider a single reactor unit equipped with protections having a predictive model constructed on a probability space that is tailored to 
	incorporate the following features.
	\begin{features}
	\end{features}
	\begin{itemize}
		\item The nuclear reactor unit under investigation is equipped with protections designed for the purpose of mitigating operational anomalies 
		and exogenous influences, called initiating events, that might lead to core damage.
		\item The sequence of initiating events occur randomly and thus forms {a} stochastic point process.
		\item A single reactor can experience  {multiple} core damage events over the course of its 
		deployment.\footnote{This modeling feature is essential in order to accommodate estimating a core damage `frequency.'}
		\item The stream of arriving initiating events is bifurcated according to the state of the protections at the time of arrival. 
		Thus, core damage events occur at epochs of those initiating events that breach reactor protections.	
	\end{itemize}
	In our arguments that follow, we employ the standard analytical framework of filtered probability spaces for characterizing `historical events' on 
	which stochastic process temporal dynamics depend. Our notation closely follows that of most standard stochastic process textbooks, 
	such as \cite{cinlar2011} or \cite{rogers2000}.
	All random processes we model are adapted to the filtration $\{\mathcal{F}_t\}_{t \ge 0}$ 
	on the filtered probability space $(\Omega, \mathcal{F}, \{\mathcal{F}_t\}_{t \ge 0}, P)$.  

	\subsection{Almost Sure Behavior}

		Let $N^D_t$, $t \ge 0$ be the number of core damage events the reactor suffers in the interval $[0,t]$.  
		\ac{cdf} is understood to be the limiting number of core damage events per unit time.
		That is, 
		\begin{definition}[Core Damage Frequency]\label{cdf_definition}
			\begin{equation}\label{c_d_f}
				\ac{cdf} = \lim_{t \rightarrow \infty} \frac{1}{t} N^D_t,  
			\end{equation}
			whenever convergence to a constant occurs \ac{as}
		\end{definition}
		\noindent \ac{cdf} is reported as a numerical constant. 
		Of course, convergence of \cref{c_d_f} is not guaranteed.  
		Further, even when \cref{c_d_f} convergences, there is no guarantee that its limit is a constant; convergence to a 
		random variable is a completely plausible circumstance.  
		We emphasize that \ac{cdf} is a numerical constant, estimates of which are used to gauge the risk of suffering a core 
		damage event.

		As will be shown, the existence of \ac{cdf} is predicated in part on the dynamics of initiating event arrivals. 
 		Let $N_t$ be the number of initiating events arriving in the interval $[0,t]$.  
		Without loss of generality we require that for almost all $\omega \in \Omega$, the trajectory $N_t(\omega)$ is right continuous and proceeds 
		in jumps of magnitude one. 
		The limiting arrival rate of initiating events $\lambda$ is defined as follows.
		\begin{definition}[Initiating Event Frequency]\label{lambda_definition}
			\begin{equation}\label{lambda}
				\lambda = \lim_{t \rightarrow \infty} \frac{1}{t} N_t,  
			\end{equation}
			whenever convergence occurs \ac{as} In general $\lambda$ can be a random variable, and convergence to a constant 
			occurs only when $\lambda$ = $E[\lambda]$
		\end{definition}
		Note that $\{N^D_t\}_{t \ge 0}$ inherits right--continuity from  $\{N_t\}_{t \ge 0}$, and it now follows directly from 
		\Cref{cdf_definition,lambda_definition} that when $\lambda$ is a constant,
		\begin{equation}\label{cdf_lambda}
			\ac{cdf} \overset{\text{\ac{as}}}= \lim_{t \rightarrow \infty} \frac{N^D_t}{{N_t}}\frac{N_t}{t}  = p \lambda
		\end{equation}
		where,
		\begin{equation*}
			p =  \lim_{t \rightarrow \infty} \frac{N^D_t}{{N_t}}.
		\end{equation*}
		Thus, \ac{cdf} is the product of the initiating event frequency with the proportion of initiating events that exceed to core damage.   
		Typically, estimating $\lambda$ is straightforward. Estimating $p$ is more challenging. 
		Note that while $p$ takes values in the interval $[0,1]$, it is defined as the limiting value of a ratio of random variables. 
		However, under certain circumstances $p$ can be interpreted as the probability that an arriving initiating event will 
		exceed to core damage.  
		One such circumstance occurs when initiating events form an ordinary Poisson process.  
		Then, the  {well-known} \ac{pasta} result applies, and $p$ can be computed as the limiting unavailability of reactor 
		protections \cite[see][]{wolff82}.  
		Unfortunately, the assumptions needed to justify \ac{pasta} defy practical justification.  
		Consequently, estimating $p$ is quite difficult in practice.

		There are a variety of approaches for crafting estimators for $p$ that incorporate the joint histories of initiating event 
		arrivals, protection maintenance activity, environmental conditions, \emph{etc.} 
		Monte Carlo methods, owing to their adaptability to complex engineering models, have gained acceptance and popularity 
		for estimating  $p$ and other statistics. 
		These methods are not, however, a panacea  {because} they require characterizing probability laws on subordinate 
		stochastic processes that must be mapped into the dynamics of protection availability in order to build useful estimators.  
		Characterizing probability laws on stochastic processes is often impractical due to the intensive data support required 
		for all but the most stylized processes.

		In order to better appreciate the manner in which \ac{cdf} jointly depends on the arrival of initiating events and the 
		efficacy of reactor protections, we will appeal to the martingale characterization of stochastic point processes~\citep[see][]{bremaud1981}.
		Let $N_t$ be (as  {stated} previously) the number of initiating events arriving in the interval $[0, t]$. 
		It is reasonable in practice to require that initiating events occur one at a time \ac{as} 
		It follows that the trajectories of $\{N_t\}_{t \ge 0}$ are nonnegative non-decreasing and proceed in jumps of size one \ac{as} 
		We define $\{X_t\}_{t \ge 0}$ as the state of reactor protections at time $t$ taking values in the set $\{0, 1\}$. 
		When an initiating event arrives to find $X_t = 1$, protection holds. 
		Otherwise, with $X_t = 0$, protection fails and core damage will ensue.

 		Both $\{N_t\}_{t \ge 0}$ and $\{X_t\}_{t \ge 0}$ are adapted to the filtration $\{\mathcal{F}_t\}_{t \ge 0}$.  
		We define $X_t$ to be left continuous and thus $\mathcal{F}_t$-predictable.  
		Clearly,  $\{N_t\}_{t \ge 0}$ forms an $\mathcal{F}$-sub--martingale. 
 		Appealing to the standard results from the martingale calculus \cite[see][]{cinlar2011}, it follows from the Doob--Mayer 
		Decomposition Theorem that
		\begin{equation}\label{Doob-Mayer}
			N_t = M_t + \Lambda_t,
		\end{equation}
		where the process $\{M_t\}_{t \ge 0}$ forms an $\mathcal{F}_t$-martingale with compensator $\{\Lambda_t\}_{t \ge 0}$ 
		($\Lambda_t$ is increasing \ac{as} and $\mathcal{F}_t$-predictable), with
		\begin{equation}\label{intensity}
			\Lambda_t = \int_0^t \lambda_s ds.
		\end{equation}
		Here, $\lambda_t$ is well defined when for all nonnegative, $\{\mathcal{F}_t\}$-predictable $\{C_t\}_{t \ge 0}$ such that
		\begin{equation}\label{martingale_lambda_definition}
			E\Big[\int_0^\infty C_t dN_t \Big] = E\Big[\int_0^\infty C_t \lambda_t dt \Big] {.}
		\end{equation}
		When well defined, $\lambda_t$ is a unique Radon--Nikodym derivative defined on the usual equivalence class, with 
		the stochastic intensity process $\{\lambda_t\}_{t \ge 0}$  {adapted} to $\{\mathcal{F}_t\}_{t \ge 0}$ and predictable. 
		Informally, $\lambda_t = E[dN_t | \mathcal{F}_{t-}]$ and can be understood as the propensity for an initiating event to arrive 
		in the next instant of time given the history of initiating events and reactor protections.

		Now, consider the martingale transform $M^D_t$ of protection unavailability $(1 - X_t)$ with respect to $M_t$ of 
		\cref{Doob-Mayer}, where
		\begin{equation*}
			M^D_t \overset{\text{def}}= \int_0^t (1 - X_s) dM_s = \int_0^t (1- X_s) dN_s - \int_0^t (1 - X_s) d\Lambda_s.
		\end{equation*}
		\begin{proposition}[Core Damage Martingale]
			$\{M^D_t\}_{t \ge 0}$ is a martingale whenever the stochastic intensity process of arriving initiating events 
			$\{\lambda_t\}_{t \ge 0}$ exists.
			\begin{proof}
				Since $X_t$ is $\mathcal{F}_t$-predictable and $0 \le X_t(\omega) \le 1$ for all $\omega \in \Omega$ 
				and $t \ge 0$, it follows that $\{M^D_t\}_{t \ge 0}$ is also an $\mathcal{F}_t$-martingale \citep[see][]{rogers2000}, and noting 
				that $N^D_t =  \int_0^t (1 - X_s) dN_s$  counts the number of core damage events in the interval $[0,t]$, 
				we have that
				\begin{equation*}\label{core_damage_martingale-0}
					M^D_t = N^D_t - \int_0^t (1 - X_s) d\Lambda_s
				\end{equation*}
				and, substituting from \cref{intensity} gives

				\begin{equation}\label{core_damage_martingale}
					M^D_t = N^D_t - \Lambda^D_t.
				\end{equation}
			\end{proof}
		\end{proposition}
		\noindent We refer to $\{M^D_t\}_{t \ge 0}$ as the \emph{Core Damage Martingale} and its compensator is given by
		\begin{equation*}
			\Lambda^D_t = \int_0^t \lambda^D_s ds = \int_0^t (1- X_s) \lambda_s ds.
		\end{equation*} 
		\begin{remark}
			\Cref{core_damage_martingale} stands as the most general expression characterizing the relationship among 
			core damage events, the arrival of initiating events, and the efficacy of reactor protections. 
			It is important to keep in mind that, for all $t \ge 0$, $M^D_t$, $N^D_t$, $\lambda^D_t$, $X_t$, and (in particular) 
			$\lambda_t$ are random variables.  
			Hence, \cref{core_damage_martingale} is nontrivial as it requires stochastic integration.
		\end{remark}

%%%%%%%%%%%%%
		Consider now, the following proposition:
		\begin{proposition}[Existence of \ac{cdf}]\label{strong_law}
			If 
			\begin{equation}
	 			\lim_{t \rightarrow \infty} \frac{M^D_t}{t} \overset{\text{\ac{as}}}= 0,
			\end{equation}
			then $\lambda^D$ exists (and is possibly a random variable) and for \ac{aa} $\omega \in \Omega$
			\begin{equation*}
				\lambda^D(\omega) = \lim_{t \rightarrow \infty} \frac{N^D_t(\omega)}{t} =  \lim_{t \rightarrow \infty} 
				\frac{1}{t} \int_0^t (1 - X_s(\omega)) \lambda_s(\omega)ds
			\end{equation*}
			where, $0 < \lambda^D(\omega) < \infty$.

			That is, 
			\begin{equation}\label{lambda^D}
				\begin{split}
					&\lim_{t \rightarrow \infty} \frac{M^D_t}{t} \overset{\text{\ac{as}}}= 0  \mbox{ if and only if} \\
					&\lim_{t \rightarrow \infty} \frac{N^D_t}{t} \overset{\text{\ac{as}}}= \lambda^D \mbox{  and  } \\
					&\lim_{t \rightarrow \infty} \frac{1}{t} \int_0^t (1 - X_s)\lambda_sds \overset{\text{\ac{as}}} = \lambda^D.
				\end{split}
			\end{equation}
			\begin{proof}
 				\Cref{strong_law} is an obvious consequence of \Cref{cdf_definition} and \cref{core_damage_martingale}.
			\end{proof}
		\end{proposition}

		\begin{remark}
			Clearly, \ac{cdf} exists only if $\frac{M^D_t}{t} \overset{\text{\ac{as}}}\rightarrow 0$ and $\lambda^D 
			\overset{\text{\ac{as}}}= E[\lambda^D ] < \infty$.  
			\Cref{strong_law} reveals the challenge in estimating \ac{cdf}.  
			In the absence of observed core damage events, predictive estimates of \ac{cdf} 
			must be formulated in terms of phenomena that can be observed. 
			To this end, analysts must rely on observations of initiating event arrival times, and reactor protection performance 
			(principally in the form of maintenance records and failure data).  
			These observations are, of course, insufficient to  capture the joint dynamics of $\{(X_t, \lambda_t)\}_{t \ge 0}$ needed 
			to directly employ the strong law relationship of \Cref{strong_law}, where,
			\begin{equation*}
				\lambda^D \overset{\text{\ac{as}}}= \lim_{t \rightarrow \infty} \frac{\Lambda^D_t}{t} = \lim_{t \rightarrow \infty} 
				\frac{1}{t} \int_0^t (1 - X_s) \lambda_s ds.
			\end{equation*}
			Monte Carlo methods do not escape the difficulty of computing $\lambda^D$.  $\{X_t\}_{t \ge 0}$ and  $\{\lambda_t\}_{t \ge 0}$ 
			are not mutually independent (even when $\{N_t\}_{t \ge 0}$ is Poisson with rate $\lambda$) and $\lambda_t$ is not directly 
			observable.  
			Since this dependence requires a Monte Carlo model to rely on an accurate characterization of the probability law on the 
			joint stochastic process $\{(X_t, \lambda_t)\}_{t \ge 0}$, it is clear that the data requirements to support accurate estimation 
			of this probability law are beyond the practical reality of reactor unit operations records.
		\end{remark}

	\subsection{Behavior in Expectation}\label{moments}

		Important insights regarding \ac{cdf} are revealed by exploring the expectation of $\frac{M^D_t}{t}$.  
		In particular, we are interested in the consequences of the stochastic dependence the state of system protections 
		$\{X_t\}_{t \ge 0}$ and the arrival of initiating events $\{N_t\}_{t \ge 0}$ and consequently $\{\lambda_t\}_{t \ge 0}$. 
		Consider now,
		\begin{proposition}[Moment Convergence]\label{moment_convergence}
			Suppose that $\frac{M^D_t}{t} \overset{\text{\ac{as}}}= 0$, then
			\begin{equation}\label{expected_strong_law}
				\begin{split}
					&\lim_{t \rightarrow \infty} E\Big[ \frac{M^D_t}{t} \Big] = 0  \mbox{ if and only if} \\
					&\lim_{t \rightarrow \infty} E\Big[ \frac{N^D_t]}{t} \Big] = E[\lambda^D] \mbox{  and  } \\
					&\lim_{t \rightarrow \infty} E\Big[ \frac{1}{t} \int_0^t (1 - X_s)\lambda_sds \Big] = E[\lambda^D].
				\end{split}
			\end{equation}
			And, with the additional condition that $\lambda^D = E[\lambda^D ] < \infty$,
			\begin{equation}\label{expected_cdf}
				\ac{cdf}= \lim_{t \rightarrow \infty} \frac{1}{t} \int_0^t E[(1-X_s) \lambda_s]ds.
			\end{equation}
			\begin{proof}
				Recall that almost sure convergence implies convergence in expectation~\cite[see][]{dudley2018}.  
				Hence, \cref{expected_strong_law} follows directly from \Cref{strong_law}.  
				Nonnegativity of the integrand in \cref{expected_strong_law} allows a routine application of Tonelli's Theorem 
				to exchange the order of expectation and integration to show \cref{expected_cdf}~\cite[see][]{folland1999}.  
				Finally, $\lambda^D = E[\lambda^D ] < \infty$ implies that 
				$\frac{N^D_t}{t} \overset{\text{\ac{as}}}\rightarrow E[\lambda^D ]$, a finite constant, thus ensuring the existence of \ac{cdf}.
			\end{proof}
		\end{proposition}
		Recalling the definition of covariance\footnote{For two random variables $Y$ and $Z$, $cov(Y, Z) \overset{\text{def}}= E[ZY] -E[Z]E[Y]$.}, 
		it immediately follows that
		\begin{corollary}\label{cov}
			When $\lambda^D = E[\lambda^D ] < \infty$, then
			\begin{equation}
				\ac{cdf} \overset{\text{\ac{as}}}= \lim_{t \rightarrow \infty} \frac{1}{t} \int_0^t cov((1-X_s), \lambda_s) ds  
				+ \lim_{t \rightarrow \infty} \frac{1}{t} \int_0^t E[(1 - X_s)] E[ \lambda_s]ds
			\end{equation}
			\begin{proof}
				Simply apply the definitions of \ac{cdf} and covariance.
			\end{proof}
		\end{corollary}
		\begin{corollary}
			When $\lambda^D = E[\lambda^D ] < \infty$,
			\begin{equation}\label{cdf_unbiased}
				\ac{cdf} \overset{\text{\ac{as}}}= \lim_{t \rightarrow \infty} \frac{1}{t} \int_0^t E[(1 - X_s)] E[ \lambda_s]ds
			\end{equation}
			if and only if $E[\lambda_t | \mathcal{F}_t] = E[\lambda_t]$ for all $t \ge 0$.
			\begin{proof}
				We need only show that $cov((1-X_t), \lambda_t) = 0$ if and only if $E[\lambda_t | \mathcal{F}_t] = E[\lambda_t]$.
		
				First, assume that $E[\lambda_t | \mathcal{F}_t] = E[\lambda_t]$ for all $t \ge 0$ for all $t \ge 0$. Note that 
				\begin{equation*}
					\begin{split}
						E[(1-X_t) \lambda_t] &= E[E[(1-X_t) \lambda_t | \mathcal{F}_t] \\
						&= E[E[(1-X_t) E[\lambda_t] | \mathcal{F}_t]\\
						&= E[\lambda_t]E[E[(1-X_t)| \mathcal{F}_t]\\
						&= E[\lambda_t] E[(1-X_t)].
					\end{split}
				\end{equation*}
				Thus, it follows from the definition of covariance  {that $cov((1-X_t), \lambda_t) = 0$}.
		
				Now, assume that $cov((1-X_t), \lambda_t) = 0$.  It follow trivially that 
				\begin{equation*}
					E[(1-X_t) \lambda_t] = E[\lambda_t] E[(1-X_t)].
				\end{equation*}
			\end{proof}
		\end{corollary}

		\begin{corollary}[Rasmussen Characterization of CDF]\label{asta}
			When, $E[\lambda_t | \mathcal{F}_t] = E[\lambda_t] = \lambda < \infty$ and $X_t \overset{\text{\ac{as}}}\rightarrow X$, 
			then
			\begin{equation}\label{pra_cdf}
				\ac{cdf} \overset{\text{\ac{as}}}= \lambda  \lim_{t \rightarrow \infty} \frac{1}{t} \int_0^t(1 - X_s)ds.
			\end{equation}
			\begin{proof}
			 	It follows from \cref{cdf_unbiased} that when $E[\lambda_t | \mathcal{F}_t] = E[\lambda_t] = \lambda < \infty$,
				\begin{equation*}
					\ac{cdf} \overset{\text{\ac{as}}}=  \lambda \lim_{t \rightarrow \infty}\frac{1}{t} \int_0^t E[(1 - X_s)]ds,
				\end{equation*}
				and since $X_t \overset{\text{\ac{as}}}\rightarrow X$,
				\begin{equation*}
					\lambda \lim_{t \rightarrow \infty}\frac{1}{t} \int_0^t E[(1 - X_s)]ds = \lambda \lim_{t \rightarrow \infty}\frac{1}{t} \int_0^t 
					(1 - X_s)ds,
				\end{equation*}
				\cref{pra_cdf} follows.
			\end{proof}
		\end{corollary}
		\begin{remark}
			It is important to appreciate that \Cref{asta} \emph{does not} imply that the state of reactor protections $\{X_t\}_{t \ge 0}$ 
			is independent of initiating event arrivals $\{N_t\}_{t \ge 0}$.  
			The condition $E[\lambda_t | \mathcal{F}_t] = \lambda$ implies that $X_t$, the state of system protections 
			at time $t$, does not influence the arrival times of future initiating events (a circumstance that we discuss more 
			fully in \Cref{conclusions}).  
			The conditions establishing \Cref{asta} allow for the possibility that initiating events can cause a failure of system protections 
			(in addition to the possibility protections were already failed immediately prior to arrival).
			\Cref{pra_cdf} has, as a special case, the condition that arriving initiating events $\{N_t\}_{t \ge 0}$ form an ordinary 
			Poisson process of rate $\lambda$. 
			We also could have shown \cref{pra_cdf} as a special case of \cref{lambda^D} where, the constant $\lambda$ can be 
			moved outside the integral \cite[see][]{wolff82} for a full development of when Poisson arrival see time averages.  
			We point out, however, that the condition $E[\lambda_t | \mathcal{F}_t] = \lambda$ can hold for initiating event streams 
			that are not Poisson \cite[see][]{melamed1990}. 
			Hence, \cref{pra_cdf} is a bit more general than the \cite{wolff82} result. 
			Of course, \Cref{asta} also holds under the more restrictive condition that  $\{X_t\}_{t \ge 0}$ and $\{N_t\}_{t \ge 0}$ are 
			independent processes.
		\end{remark}

\section{Summary and Conclusions}\label{conclusions}

	The appeal of \ac{cdf} as a risk metric is that it reports as a single numerical value that is computed using historical data captured 
	for a given reactor unit using physics and operations-based predictive modeling.
	The magnitude of the reported value is intended to quantify expectations regarding time between core damage events.  
	That is, the reciprocal of \ac{cdf} estimates the ``expected'' time between core damage events.  
	And, since in practice, an operating reactor unit has never suffered a core damage event, \ac{cdf} is intended to offer insight as to expectation of the 
	remaining time until the first core damage event might occur.

	Estimates of \ac{cdf} are almost always developed through \ac{pra} and \ac{psa} studies.  
	These studies use combinational logic models (fault trees) of reactor protections subordinated to a Monte Carlo simulation experiment. 
	Monte Carlo relies on estimates of the probability law for the various fault tree elements and initiating events.  
	Fault tree element probability laws (\emph{i.e.,} reliability functions) are estimated from historical data associated with reactor unit operations and maintenance. 
	Initiating event laws are developed from histories of both exogenous influences (\emph{e.g.,} weather events, seismic disturbances, \emph{etc.}) 
	and endogenous influences (\emph{e.g.,} equipment failures, human operator errors, \emph{etc.}).  
	Modeling reliance on historical data reveals the necessity of strong law characterizations of event probabilities.  
 	But, it is important to appreciate that Level-1 \ac{pra} studies rely on the Rasmussen characterization of \ac{cdf}. 
	That is, either through explicit {or} implicit assumption, future arriving initiating events are taken to be independent of the present 
	state of reactor protections. 
	Consequently, Monte Carlo estimates of \ac{cdf} developed through  {\ac{pra} converge} to the form given in \cref{cdf_unbiased}.  
%
%We will comment further in \Cref{underestimate} on why escaping the Rasmussen characterization and its consequences are nearly impossible in practice.

	Whether or not the appeal of \ac{cdf} as a risk metric is justified is the subject of much literature that we will not review. 
	It is nonetheless fact that \ac{cdf} is a quantity of interest to regulators, design engineers, utility operators, and the public, and is a focus 
	of Level-1 \ac{pra} studies. 
	Our focus is limited to the existence of \ac{cdf} and the fidelity of its estimates.  
	To this end, we revisit the three question posed in \Cref{introduction}.

	\subsection{When does \ac{cdf} exist?}

		We establish the conditions required for the existence of \ac{cdf} in \Cref{strong_law}. 
		We observe that existence relies on the strong convergence to zero of the core damage martingale.  
		As \Cref{strong_law} shows, convergence cannot be achieved when either reactor protections or initiating events do not achieve 
		a steady state regime. 
		For instance, if climate change is deemed to be in play during the operational life time of a reactor unit, then exogenously arriving 
		weather--response initiating events are unlikely to be in steady state.  
		Similarly, our present understanding of seismic activity relies on self--exciting stochastic point processes.  
		This suggests that exogenous initiating events arising from seismic activity are unlikely to achieve steady state during a reactor 
		unit's operating lifetime.  
		Finally, and perhaps most importantly, much engineering effort is devoted to discovering and eliminating reactor protection failure 
		modes not included in preliminary \ac{fmea} studies.  
		These discoveries and corresponding design corrections dramatically influence the probability law on $\{X_t\}_{t \ge 0}$ and 
		whether or not reactor protections converge to a steady state regime. 
		It is important recognize that the discovery of any new failure mode after time $t$ cannot appear in the prior history captured in the 
		filtration $\{\mathcal{F}_t\}_{t \ge 0}$; this practical engineering reality suggests that ``good engineering practice'' and innovation contribute to the 
		lack of stationarity for reactor protections.

	\subsection{Is it feasible to estimate \ac{cdf}, when it exists?}\label{feasible}

		Accepting the existence of \ac{cdf}, it is a straightforward matter to compute estimates under the Rasmussen characterization; 
		this is the function of Level-1 \ac{pra} studies. 
		If, however, the Rasmussen characterization is rejected, \ac{cdf} is not easily estimated.  
		In order to better understand this, let $T_n$, $n \ge 1$ be the time of the $n^{th}$ arriving initiating event.
		\begin{equation*}
			T_n = \inf \{t > 0: N_t \ge n\}, n \ge 1.
		\end{equation*}
		When \ac{cdf} exists,
		\begin{equation}\label{def_T_n}
			X_{T_n} \overset{\text{\ac{as}}}{\rightarrow} \tilde{X} \mbox{ and } X_t \overset{\text{\ac{as}}}{\rightarrow} X.
		\end{equation}
		Under the Rasmussen characterization,
		\begin{equation}\label{rasmussen}
			\tilde{X} \overset{\text{\ac{as}}}= X \implies E[\tilde{X}] = E[X] \overset{\text{\ac{as}}}= \lim_{t \rightarrow \infty} \frac{1}{t} \int_0^t X_s ds.
		\end{equation}
		That is, the Rasmussen characterization requires that the state of reactor protections as seen at epochs of initiating event arrivals is 
		the same as would be seen at an arbitrary time.

		Of course, relaxing the requirements of the Rasmussen characterization implies that $E[\lambda_t | \mathcal{F}_t] \ne \lambda$ which has the consequence that 
		\cref{rasmussen} does not hold.  
		\Cref{cov} reveals the bias between true \ac{cdf} and the Ramussen characterization as
		\begin{equation}\label{bias}
			\ac{cdf} - \lambda (1 - E[X]) \overset{\text{\ac{as}}}=  \lim_{t \rightarrow \infty} \frac{1}{t} \int_0^t cov((1 - X_s), \lambda_s) \ne 0,
		\end{equation}
		and while the bias is  {known} to exist, stochastic intensity $\lambda_t$ is a Radon--Nikodym derivative that is not directly observable; 
		hence, there are no data to support a Bayesian estimator either parametric or non--parametric.

	\subsection{Do estimates of \ac{cdf} closely approximate its true value?}\label{underestimate}
	
		Any special properties associated with the bias given \cref{bias} are of considerable practical interest. 
		To this end, recall the definition of the Pearson correlation coefficient $\rho(Y, Z)$ between any two random variables $Y$ and $Z$ defined on 
		a common probability space where
		\begin{equation*}
			\rho(Y,Z) \overset{\text{def}}= \frac{cov(Y,Z)}{\sqrt{var(Y)} \sqrt{var(Z)}}.
		\end{equation*}
		$\rho(Y,Z)$  is positive if and only if Y and Z tend to be simultaneously greater than, or simultaneously less than, their respective expected 
		values.\footnote{The converse is true for a negative value of $\rho(Y,Z)$.}

		It immediately follows from the definition of correlation coefficient that $(1-X_t)$ and $\lambda_t$ are either positively correlated or uncorrelated 
		if and only if $cov((1-X_t), \lambda_t) \ge 0$.  
		Observe that for each $t, s \ge 0$,  and $\omega \in \Omega$, $X_{t+s}(\omega) - X_t(\omega)$ takes values only the the set $\{-1, 0, 1\}$.

		Straightforward engineering reasoning reveals that for all $t, s > 0$,
		\begin{equation*}
			\lim_{s \downarrow 0}P(X_{t+s} - X_t = 1, N_{t} - N_{t-s} = 1) = 0.  
		\end{equation*}
		This is to say, \emph{almost surely an arriving initiating event will not cause non-functioning reactor protections to suddenly function}, 
		and it follows that
		\begin{equation*}
			\lim_{s \downarrow 0}P(X_{t+s} - X_t \le 0 | N_{t} - N_{t-s} = 1) = 1.
		\end{equation*}
		This is equivalent to saying that $X_t$ almost surely will not increase given an increase in $N_t$.  
		Since $N_t$ is nondecreasing in $t$, we observe that $X_t$ and $N_t$ must be either negatively correlated or uncorrelated. 
		As an immediate consequence, $(1-X_t)$ and $N_t$ must be either positively correlated or uncorrelated.  
		Noting the informal relationship that $E[dN_t| \mathcal{F}_{t-}] = \lambda_t dt$, it is easily reasoned that $\lambda_t$ and $(1-X_t)$ must be either positively 
		correlated or uncorrelated.  
		%
		%Positive correlation between $\lambda_t$ and $(1-X_t)$ is a direct consequence of relaxing the assumption of non--damage 
		%and considering the consequences of arriving initiating events disrupting protections that might otherwise have suppressed core damage. 
		%
		This circumstance agrees with engineering experience.\footnote{For example, when 12 of 13 emergency backup 
		diesel generators were washed away at Fukushima Daiichi, loss of onsite power and thus the ability to sustain cooling lead to core 
		meltdown in each of the reactor units.}
		Thus, when $\lambda_t$ and $(1 - X_t)$ are positively correlated or uncorrelated, increasing the propensity for an initiating event 
		to arise in the next instant of time increases the propensity for core damage.
		Similarly, decreasing the propensity of an arriving initiating event in 
		the next instant of time decreases the propensity for core damage. 
		More precisely,
		\begin{equation*}
			\lim_{s \downarrow 0}P(X_{t+s} - X_t \le 0 | \tilde{\lambda}_t) \overset{\text{\ac{as}}}> \lim_{s \downarrow 0}P(X_{t+s} - X_t 
			\le 0 | \hat{\lambda}_t), \mbox{ for any }\tilde{\lambda}_t \overset{\text{\ac{as}}}> \hat{\lambda}_t
		\end{equation*}
		and
		\begin{equation*}
			\lim_{s \downarrow 0}P(X_{t+s} - X_t \le 0 | \tilde{\lambda}_t) \overset{\text{\ac{as}}}< \lim_{s \downarrow 0}P(X_{t+s} - X_t \le 0 | 
			\hat{\lambda}_t), \mbox{ for any }\tilde{\lambda}_t \overset{\text{\ac{as}}}< \hat{\lambda}_t.
		\end{equation*}
		We, thus, conclude that since for all $t \ge 0$,  $(1- X_t)$ and $\lambda_t$ cannot be negatively correlated, the bias of \cref{bias} 
		must be nonnegative.  
		The nonnegative bias appearing in \Cref{cov} ensures that when \ac{cdf} exists, the Rasmussen characterization always 
		underestimates the true value of \ac{cdf}.

		In practice, numerical values for \ac{cdf} are typically computed within a \ac{pra} or \ac{psa} study that relies on the Rasmussen characterization for Level-1 studies.
		The inadequacy of the Rasmussen characterization, of course, derives from how (typically very sophisticated) protection (un-)availability analysis in incorporated into 		\ac{pra}.  Rasmussen insightfully seeks to extended traditional protection availability analysis, by noting that the risk of a core damage event should be understood as 		a consequence of an unmitigated initiating event.  Hence, it is only at arrival times of initiating events that protection unavailability would lead to a core damage event. 		Treating $\lim_{t \rightarrow \infty}(1-X_t)$ as the top event of a fault tree analysis of the unit protective system, the Rasmussen characterization reasons that $p = 			\lim_{t \rightarrow \infty} E[(1 - X_t)]$ is the likelihood of core damage at the time of an initiating event. However, a closer examination reveals that unavailability at 			instants of initiating events should be written as $\lim_{n \rightarrow \infty} E[(1- X_{T_n})]$.  Since the index on unavailability at the time of initiating event arrivals is 		the sequence of random variables $\{T_n; n\ge 1\}$, it is logical that, except under very special conditions,  $\lim_{n \rightarrow \infty} E[(1- X_{T_n})] \neq \lim_{t 			\rightarrow \infty} E[(1- X_{t})]$.
		
		An arriving initiating event will find system protections either working or failed. If the protections are working, they will ensure that the initiating event does not lead to 		catastrophe.  On the other hand, if protections are failed upon arrival, catastrophe will ensue.  Catastrophe ensues in two distinct scenarios:
		    \begin{enumerate}
		        \item System protections are failed at time $T_n$ for reasons unrelated to the arrival of the $n^{th}$ event.
		        \item System protections were working, but the arrival of the $n^{th}$ event caused protection failure.
		    \end{enumerate}
		Clearly, the covariance bias term in \Cref{cov} is associated with item 2.  Because the Rasmussen characterization relies on protective system reliability 					modeling which only describes $(1-X_t)$, $t \ge 0$, uncertainty associated with the initiating event arrival time point process  $\{T_n; n \ge 0\}$ is not represented.
		Upon recognizing this modeling deficiency, it might seem reasonable to extend the Rasmussen characterization so as to overcome its optimistic bias.  But, the very purpose of exploring \ac{cdf} is to approximate it with a numerical value that is supported by operations and maintenance date.  Yet, there will exist not observational data to support estimating protection unavailability $(1 - X_{T_n})$, for any $n \ge 1$ because in practice we have no historical core damage events In practice, an unbiased estimate of \ac{cdf}, in the absence of any observed core damage events should (except under very special circumstances) seem too good to be true, whenever initiating events are believed to effect protection reliability.  Rasmussen and his successor \ac{pra} researchers were ensnared by a common paradox (the random observer problem) that is untangled in the large and well--known Palm Calculus literature.  Unfortunately, understanding the Rasmussen bias is not the same as escaping it.  There is no ``fix'' to correct this bias, and this deficiency has important implications for \ac{pra} and \ac{psa} and risk-informed regulatory oversight.

	\subsection{Final Observations}
		 
		We have shown that the Rasmussen characterization of \ac{cdf} will at best yield values that underestimate true \ac{cdf} and at worst 
		provide values for a metric that does not exist. We observe from corollary \ref{cov} of proposition \ref{moment_convergence} that the optimistic bias of \ac{cdf} occurs under the Rasmussen characterization because it generally does not account for the possibility that an arriving initiating event exceeds to catastrophe by damaging the functionality of system protections.  
		Clearly, it is infeasible to quantify such catastrophes since historical data does not exist. 
		Further, reliance on \emph{ad hoc} characterizations of the Rasmussen characterization bias lack the rigor necessary to support the objectives of high--consequence risk 
		analyses like \ac{pra} and \ac{psa}.
		%
		%This is to say, there is no straightforward path for extending or modifying \ac{pra} or \ac{psa} to quantify bias.  
		%
		Thus, philosophical objections regarding the usefulness of \ac{cdf} aside, reported values of \ac{cdf} are not particularly 
		informative as a risk metric quantifying reactor safety because their optimistic bias cannot be analytically quantified.  This observation calls into question the use of \ac{cdf} in regulatory language.  The intent of regulators oversight is to ensure adequate safety for the public.  To this end, the public is better served by well--developed prescriptive regulations than optimistically biased risk--informed regulations built upon \ac{pra} or \ac{psa} analyses.

\bibliographystyle{chicago}
\bibliography{New_CDF}

\end{document}